\documentclass[a4paper]{jhep3}

\usepackage{amsmath}
\usepackage{amsfonts}
\usepackage{bbm}

\title{On the distance observable in the Moyal plane
and in a novel two-dimensional space with string-theory pregeometry}

\author{Giovanni AMELINO-CAMELIA*, Giulia GUBITOSI and Flavio MERCATI\\
Dipartimento di Fisica, Universit\`a di Roma ``La Sapienza" \\
and Sez.~Roma1 INFN \\P.le A. Moro 2, 00185 Roma , Italy\\ *Email:
\email{Giovanni.Amelino-Camelia@roma1.infn.it}}

\abstract{Motivation for the study of spacetime noncommutativity comes primarily
from its possible use in investigations of (Planck-scale) spacetime fuzziness,
but most work focuses on S-matrix/field-theory observables
and still very little has been established for geometric observables.
We argue that it might be useful to
exploit the "pregeometric" formulation of
spacetime noncommutativity, which in particular describes the coordinates
of the Moyal plane in terms of the phase-space coordinates of a point
particle "living" in a auxiliary/fictitious spacetime
($\{x_1,x_2\}_{Moyal} \equiv \{q,p\}_{particle}$).
This leads us straightforwardly
to a distance operator for the Moyal plane,  and allows us to
expose some limitations of a previous attempt to describe the "area
of a disc" in the Moyal plane. We also observe that from our pregeometric
perspective it is rather natural to contemplate a spacetime whose
pregeometric picture is based on the phase-space coordinates (fields)
of a string.
The fact that such "stringspaces"
essentially provide spacetime points with extendedness is relevant for
the fuzziness of geometric spacetime observables in ways that we preliminarily characterize
through an analysis of the distance observable on a two-dimensional stringspace
and through the observation that
by implementing the Amati-Ciafaloni-Veneziano/Gross-Mende uncertainty relation in the phase
space of the pregeometric string one could have stringspaces with a novel type of
fuzzy coordinates.}

\begin{document}

\maketitle

\section{Introduction and summary}

The hypothesis that the fundamental description of spacetime structure should involve
one form or another of quantization has been explored extensively, mostly (but not exclusively)
within the literature devoted to the study of the quantum-gravity problem.
One of the most studied forms of quantization introduces noncommutativity of the
coordinates of a Minkowski-type (flat) spacetime in a sense that attempts to reproduce the
success of the Heisenberg noncommutativity of the phase-space position coordinates
of point particles.
A simple and much studied example is
the (Groenewold-)Moyal plane~\cite{moyalrefs1,moyalrefs2,moyalrefs3},
which is a two-dimensional flat space with noncommutativity
of the coordinates $x_1$ and $x_2$ given by:
\begin{equation}
\left[ x_1, x_2\right]=i \theta~.
\label{moyal}
\end{equation}
For four-dimensional spacetime noncommutativity the simplest possibility
is still a ``canonical spacetime"~\cite{moyalrefs3},
with coordinates such that ($\{ \mu,\nu \} \in \{ 0,1,2,3 \}$)
\begin{equation}
\left[ x_\mu, x_\nu \right]=i \theta_{\mu\nu}~.
\label{canonical}
\end{equation}
[The noncommutativity parameters ($\theta$ in (\ref{moyal})
and $\theta_{\mu \nu}$ in (\ref{canonical}))
 commute\footnote{While over the last few years canonical spacetimes are indeed mostly
 studied assuming that $\theta_{\mu \nu}$
 commutes with the coordinates, the research programme that first led to the proposal
 of canonical noncommutativity~\cite{dopli1994} is contemplating the possibility~\cite{doplithetanontrivi}
 of a $\theta_{\mu \nu}$ that does not commute with the coordinates.}
  with the coordinates.]

Most of the interest in these studies,
especially from the quantum-gravity side, originates primarily from heuristic
arguments suggesting~\cite{mead,padma,ngmpla,gacmpla,ahluGUP,garay}
that when gravitational and quantum-mechanical effects are both taken into account
it should not be possible to determine sharply the coordinates of a spacetime point/event.
However, the ``fuzziness" of noncommutative spacetimes has not been extensively
investigated. Instead most studies focus on the structure of field theories introduced
in such spacetimes and on ``S-matrix observables" (such as the implications of noncommutativity
for the probability of occurrence of certain particle-physics processes).

We here argue that a valuable tool for the exploration of the fuzziness
could be provided by the pregeometric perspective on spacetime noncommutativity, which
essentially is the idea of exploiting fully the correspondence between certain types
of spacetime noncommutativity and the noncommutativity of the Heisenberg phase space.
For example, certain properties of functions of the coordinates of the Moyal plane
can be obtained describing them as operators acting on the Hilbert space
of a particle ``living" in a pregeometric/auxiliary/fictitious spacetime.
The availability of such a pregeometric picture is rather obvious and has already been stressed
by several authors, but so far
its potentialities of providing intuition (and tools) for computations
that one might want to do in the noncommutative-spacetime setting have been explored
only in a very limited way.

In the next section we report our analysis of the distance operator for the Moyal plane,
for which the pregeometric perspective proves to be valuable. Interestingly
our distance operator has a discrete spectrum
and a (non-zero) minimum  eigenvalue.
While we are not aware of any previous published studies of the distance operator
in the Moyal plane, our comments on the area operator are offered (in
the short Section 3) also in relation to the analysis of
the area of a disc in the Moyal plane reported in Ref.~\cite{romero}.
We shall attempt a detailed analysis of the area-observable issue in a forthcoming paper,
but the observations we report here suffice to expose some limitations
of the proposal put forward in Ref.~\cite{romero}.

In section 4 we discuss briefly other applications
(beyond Moyal)
of the conventional pregeometric
picture for spacetime noncommutativity, and we also propose a possible new way
to make use of the pregeometry perspective:
while it usually intervenes to provide intuition on
how to derive some characterizations of a preexisting (noncommutative) spacetime geometry,
we argue that it might be useful to adopt a ``pregeometric perspective" for the task
of devising spacetime geometries of certain desired structures.

In Section 5 we give a first example of the new way to make use of
the pregeometric perspective which we advocate.
Specifically we observe that a way to devise
a spacetime geometry with coordinates each affected by irreducible uncertainties
(whereas in standard noncommutative geometries irreducible
uncertainties are found only for combinations of coordinates)
can perhaps be inspired by the string-theory Amati-Ciafaloni-Veneziano/Gross-Mende
results on uncertainty relations
for the phase space of strings: if the pregeometric picture of a point of
spacetime is given by a (functional) point in the phase-space of such strings
then the coordinates of this (``extended") spacetime points should indeed be individually
affected by irreducible uncertainties. On the technical side we are at present
not ready to develop in detail such a fully articulated string-theory-based pregeometric picture,
but we do attempt to provide a preliminary intuition for its structure through an
exploratory analysis of some properties of the distance operator in such a ``stringspace"
(for the case of a two-dimensional stringspace and
relying only on a rudimentary characterization of the string-theoretic
pregeometry).

In the closing Section~6 we offer some brief considerations on the outlook
of application of the ``pregeometric strategy" proposed in this paper
(both for the Moyal plane and other conventional noncommutative spaces, and
for our stringspaces).

\section{Pregeometry and distance operator for the Moyal plane}
\label{SezioneDistMoyal}

A standard pregeometric description
of a point in the Moyal plane is obtained in terms of the phase-space coordinates
of a fictitious nonrelativistic quantum particle ``living"
on a 1+1-dimensional pregeometric spacetime.
Denoting by $\xi$ the spatial coordinate of the pregeometric spacetime,
the states of the fictitious particle are described by ``wave functions" $\psi(\xi)$
and its phase-space ``observable'' coordinates $\hat \xi$ and $\hat \pi$ are noncommutative operators
acting on the space of $\psi(\xi)$'s, subject to the Heisenberg commutation relation
\begin{equation}
 \left[\hat \xi,\hat \pi\right]=i \hbar_{\text{pregeom}}~.\label{eq:heisenbergcommutator}
\end{equation}
Clearly one obtains a faithful characterization of the Moyal plane by
taking\footnote{In this paper we reserve the symbol ``$\sim$", which ordinarily
describes the near equality of
two quantities, to characterize the link between structures in the ``physical" space
and their pregeometric description in terms of observables in the pregeometry. }
\begin{equation}
x_1 \sim \hat \xi~,~~~x_2 \sim \hat \pi~,~~~\theta = \hbar_{\text{pregeom}} ~.\label{identification}
\end{equation}
On the basis of this description of the Moyal plane
it is possible to make use of well-established properties of the phase-space
observables of nonrelativistic quantum particles on a one-dimensional space
to characterize at least some aspects of the geometric structure of the
Moyal plane.

Note that our choice of conventions attempts to provide easy-to-recognize characterizations
of the variables that appear in our analysis. A variable with a hat (say, $\hat q$)
is always an observable (be it physical or pregeometric),
while variables without a hat (say, $q$) are always coordinates on the geometry of the spacetime
framework in which observables (such as $\hat q$) are introduced. In particular
here $\xi$ is the (algebraically trivial) spatial coordinate on the fictitious
spacetime where
the pregeometric theory is formulated, $\hat \xi$ is the pregeometric observable that
gives the position of a particle in the space with coordinates $\xi$, and $x_1 , x_2$
are (noncommutative, Moyal) spatial coordinates of the spacetime which should be considered
as the ``physical" spacetime of our picture. We would denote by $\hat x_1 , \hat x_2$
observables describing the position of a nonrelativistic particle in the Moyal plane (they should
be operators acting on some wave functions, $\psi(x_1 , x_2)$, whose arguments are noncommutative
coordinates).

Both for its intrinsic interest and because it is
a valuable illustrative example of Moyal-plane analysis
that can be based on the pregeometric perspective, we report here a study
of the distance observable in the Moyal plane.
In light of the characterization of points in the Moyal plane given above,
one should evidently be able to describe the
distance between two points of the Moyal plane
as some appropriate operator acting on two-particle pregeometric states.
Let us denote the Moyal-plane (noncommutative) coordinates
of one of the points with $x_1^{(1)} , x_2^{(1)}$
and use for the other point $x_1^{(2)} , x_2^{(2)}$.
These can indeed be described (pregeometrically) in terms of the following
operators acting on two-particle pregeometric states:
\begin{equation}
x_1^{(1)}\sim \hat \xi \otimes \mathbbm 1 ~,~~~
x_2^{(1)} \sim \hat \pi \otimes \mathbbm 1 ~,~~~
x_1^{(2)} \sim \mathbbm 1\otimes \hat \xi  ~,~~~
x_2^{(2)} \sim \mathbbm 1\otimes \hat \pi ~.
\end{equation}

As the observable (squared-)distance between two points in the Moyal plane
it is natural to take the standard formula\footnote{Besides the familiarity of its structure,
as we shall discuss elsewhere~\cite{agmNEXT}, the operator $d^2$ introduced
in (\ref{distanzamoyal}) appears to be a natural
choice also because
 it turns out to be compatible with the (rather trivial) symmetries
of the Moyal plane.}:
\begin{equation}
 d^2=\left(x_1^{(1)}-x_1^{(2)}\right)^2+\left(x_2^{(1)}-x_2^{(2)}\right)^2 ~.
 \label{distanzamoyal}
\end{equation}

Valuable insight on the properties of $d^2$ is obtained, within the pregeometric setup,
by straightforward analysis of the spectrum of a would-be-Hamiltonian operator $\hat H$
for two-particle pregeometric states. One in fact can take
(using again the symbol $\sim$ to link a structure in the Moyal space
to an observable in the pregeometry)
\begin{equation}
d^2 \sim  \hat H \equiv \frac{\hat p^2}{2 m}+\frac{1}{2}m \omega^2 \hat q^2,
\end{equation}
with $\hat q \equiv \hat \xi \otimes \mathbbm 1 - \mathbbm 1\otimes \hat \xi$,
 $\hat p \equiv \hat \pi \otimes \mathbbm 1 - \mathbbm 1\otimes \hat \pi$,
 $m=\frac{1}{2}$, $\omega=2$.
Then, also noticing that from (\ref{eq:heisenbergcommutator}) it follows
that $\left[\hat q,\hat p\right]=2i\hbar_{\text{pregeom}}=2i \theta$,
one easily concludes that the spectrum of $d^2$ is
\begin{equation}
 d^2_n = E_n = 4 \theta \left( n+\frac{1}{2} \right)~,
\end{equation}
with $n$ integer and nonnegative.

Since one of the motivations for investigating the idea of a quantum spacetime
originates from interest (see, {\it e.g.}, Ref.~\cite{areaLQG})
in the possibility of discrete spectra for spacetime
geometric observables, it is noteworthy that the (squared-)distance observable
in the Moyal plane is discrete, with separations of $4 \theta$ between the eigenvalues,
and has a minimum value: $d^2_{min}=2 \theta$.

\section{On the area observable in the Moyal plane}
We are not aware of any previous published studies of the distance observable
in the Moyal plane, and our original result concerning discretization
and a minimum value for this observable might well prove to be an insightful
characterization of Moyal noncommutativity.
Having figured out a few things on the distance observable
the next natural tasks should concern the area observable,
and in the previous literature one does find a study, in Ref.~\cite{romero},
that might be relevant for the area observable in the Moyal plane.
In a quantum-spacetime setup
the concept of area of a surface is much more challenging than the one of
distance between two points, and
we postpone a detailed discussion of the area observable to a forthcoming
paper~\cite{agmNEXT}, but we do want to offer here a few remarks
characterizing the complications that the description of the area of a surface
entails in a noncommutative space, also observing that the pioneering attempt
of analysis of areas reported in Ref.~\cite{romero} might require rather
substantial revisions.

Let us indeed start with a brief characterization of the
results reported in Ref.~\cite{romero},
which focused on an attempt to describe the area of a disc
in the Moyal plane.
A first potentially worrisome aspect of
the analysis reported in Ref.~\cite{romero} originates from the
fact that the disc is specified through a single point in the
Moyal plane, with noncommutative coordinates $ x_1, x_2$.
The area $A_{disc}$ of the disc was then
described as follows~\cite{romero}
\begin{equation}
 A_{disc}=\pi\left[ x_1^2 + x_2^2\right]. \label{eq:arearomero}
\end{equation}
in terms of the noncommutative coordinates $ x_1, x_2$
of that single point.   The analysis ends up proposing a spectrum
for $ A_{disc}$ which is discrete and has a minimum (nonzero) value.

First of all let us notice that, in a plane characterized
by uncertainty relations of the type
\begin{equation}
\delta x_1 \delta x_2 \geq \theta \label{moyalUncertain}
\end{equation}
it is to be expected that, as we found in the previous section,
the distance observable has a nonzero minimum value,
but it would be rather paradoxical to have a nonzero minimum allowed
value of areas.
A surface is identified by a network of points and it suffices to have
the case of all points with the same value of, say, $x_1$
to have then zero area.
And the requirement $\delta x_1 \delta x_2 \geq \theta$
of course does not exclude the possibility of
a network of points all with the same value of $x_1$
(the $x_2$ coordinates of these points may well be affected by
strong uncertainties, but the area would be zero exactly/sharply).
For the distance between two points to be zero instead
one should demand that all coordinates of the two points
coincide, which is very clearly incompatible
with $\delta x_1 \delta x_2 \geq \theta$, and indeed
we did report a minimum-distance result
in the previous section.

So it seems that we should conclude that the minimum-area result
of Ref.~\cite{romero} must be an artifact, the result of some unwarranted
assumptions made in the derivation. The idea of characterizing
a surface through a single point on the boundary of the surface,
as done in Ref.~\cite{romero}, should have been subject to further scrutiny
in any case, and must then be viewed even with more skepticism
 in light of the puzzling mismatch between the
derived properties of areas and the uncertainty
relations $\delta x_1 \delta x_2 \geq \theta$.

\section{Pregeometry: some standard applications and a new strategy}
In physics and mathematics the characterization as ``pregeometry" has been introduced
in many different ways, from several different perspectives. It is clearly
beyond the scopes of this paper to propose a ``historical route"
through this broad literature.
We do need to characterize our own perspective on ``pregeometry", and for this we
find useful to link it to two of the concepts of pregeometry introduced in the
physics literature.
Our first influence is perhaps the most classical concept of pregeometry
contemplated in the quantum-gravity literature, which was championed by Wheeler
already in the 1960s and 1970s.
This is the idea that perhaps a solution to the quantum-gravity problem
might be more easily reached if, rather than working in ways that refer directly
to spacetime, the formalisms developed looking for candidate solutions of the
quantum-gravity problem were articulated in terms of some deeper underlying rules of physics
from which then in the end spacetime would be in some appropriate sense derived.
Our second influence is the concept of pregeometry as sometimes used
in studies of spacetime noncommutativity, where by describing the noncommutative spacetime
coordinates in terms of Heisenberg phase-space observables of an ``underlying" (but
possibly only fictitious, mathematically auxiliary) particle theory one finds ways
to perform more intuitively certain computations that spacetime noncommutativity invites
one to contemplate.

The perspective we advocate here is roughly a hybrid of these two influences,
further elaborating on a ``pregeometric perspective" of similar hybrid nature
advocated in Ref.~\cite{gacmajid} (which in turn
was elaborating on earlier ``pregeometric perspectives" proposed by Majid~\cite{majidPregeom}).
The ways in which we would contemplate formalizing the pregeometric level make use
of particle-physics (and string-theory, see later) intuition, in a sense that is closely
related to the pregeometric techniques sometimes used in the noncommutative spacetime literature.
But the physical picture (and therefore the entities introduced at the pregeometric level
and the interpretation of the symbolism used at the pregeometric level)
is guided by the physical-problem-solving spirit of the Wheeler perspective.
And, while in  the noncommutative-spacetime literature usually a formulation of the
properties of spacetime comes first and then a pregeometric description is later identified,
mainly as a convenient tool for computations,
here we see reasoning at the pregeometric level as an opportunity to devise new
pictures of spacetime structure. By limiting ourselves, in the search of
alternative ways to describe the structure of spacetime,
to the use of intuition that is indigenous to spacetime itself we might be missing
on opportunities provided by our experience with the large variety of physical
systems Nature confronts with: physics has encountered many ``spaces" along the way
to characterize the diversity of Nature and each of these spaces is in principle
a possible source of intuition for the structure of the space we most care about,
which is spacetime. Besides the example of the phase space of a particle, usually adopted
in pregeometric pictures of spacetime noncommutativity, one can (in a sense that will be
roughly illustrated in the next section) consider the phase space of a string,
and (in a sense that we hope to be in a position to illustrate in the not-so-distant future)
even spaces that at first do not appear to look  like spacetime (or even phase space)
could well come into play, and a particularly amusing possibility would be the one
of associating spacetime to some space of solutions of a dynamical system (amusing because
it would link fully dynamical entities in the pregeometry to the kinematic/structural
properties of the physical spacetime\footnote{This would open strikingly novel (and, admittedly,
perhaps ``too novel") possibilities, as illustrated by the fact that for example for world
lines in some pregeometry one could introduce the concept of braiding, while a concept
of ``braiding of points"
(points of spacetime) is something we might struggle to come to terms with (and perhaps
indeed we might prudently choose not to). }).

In a sense that perhaps is close to the one intended by Wheeler, we envision that
it might be in some cases advantageous/fruitful to adopt as laws of dynamics of spacetime
 some laws that are primitively
introduced (and most naturally viewed) at the pregeometric level, and that would amount to
spacetime dynamics only as a result of the link established between spacetime structures
and pregeometric entities.

The tentative characterization of our work in progress offered in this section is
relevant to the present paper only inasmuch as it provides some tools of interpretation
for our readers of the motivation for the observations reported in the second part of
this paper. While the observations reported in the previous two sections reflect
a type of use of pregeometry that is standard
in the noncommutative-spacetime literature, the proposal sketched out in the next section
can be perhaps more intelligible to our readers in light of the perspective
we are offering in this section.

In closing this section let us stress that, while both
in the first part and in the second part of this paper the starting point
is the Moyal plane (investigated in the first part, and generalized, in
appropriate sense,
in the second part), some of the techniques that are available in the
noncommutative-spacetime
literature should easily allow to generalize all of our results to richer
examples of spacetime noncommutativity.
For the much-studied
four-dimensional canonical spacetimes, which (as already reported in
Eq.~(\ref{canonical}))
are characterized by $[ x_\mu ,  x_\nu] = i \theta_{\mu\nu}$, an example of
pregeometric
description is obtained by posing (see, {\it e.g.}, Ref.~\cite{calmet})
$$
 x_\mu \sim  \hat \xi_\mu - \frac{1}{2 \hbar_{\text{pregeom}}} \delta^{\sigma
\rho}\theta_{\mu\sigma} \hat \pi_\rho  ~,
$$
where $\delta^{\mu\nu}$ is the Kroenecker delta , and $\hat \xi_\mu$ and $\hat
\pi_\nu$ are phase-space coordinates (observables) of a pregeometric
particle ``living"
in a 5-dimensional auxiliary spacetime:
$$
[\hat \xi_\mu , \hat \pi_\nu] = i \delta_{\mu\nu} \hbar_{\text{pregeom}},
\qquad [\hat \xi_\mu
, \hat \xi_\nu] = [\hat \pi_\mu , \hat \pi_\nu] = 0 ~.
$$

And for the four-dimensional ``$\kappa$-Minkowski spacetime", characterized by
$$
[ x_j, t]=i\lambda  x_j \quad [ x_j, x_k]=0~,
$$
on which there is also
a rather sizeable literature,
several alternative formulations in terms of
the phase spaces of particles ``living" in a four-dimensional
auxiliary/fictitious spacetime
have been contemplated, including the possibility
$$
 x_j \sim \hat \xi_j , \qquad t \sim e^{\lambda (\hat \pi_1+\hat
\pi_2+\hat \pi_3) /\hbar_{\text{pregeom}}} ,
$$
$$
[\hat \xi_j , \hat \pi_k] = i \delta_{jk} \hbar_{\text{pregeom}} .
$$
and the possibility~\cite{majidPregeom,gacmajid}
of the limit $\rho,\hbar_{\text{pregeom}}\rightarrow\infty$
(with $\frac{\hbar_{\text{pregeom}}}{\rho}=\lambda$)
of\footnote{But
in this case there would be strong mathematical
motivation~\cite{majidPregeom,gacmajid}
for introducing in the pregeometric picture also
the following nontrivial coalgebraic property: $\Delta ({\hat \xi}_j) = {\hat \xi}_j  \otimes
\mathbbm{1} + \mathbbm{1} \otimes {\hat \xi}_j ,\quad
 \Delta ({\hat \pi}_j) = {\hat \pi}_j \otimes \mathbbm{1} + e^{-\frac{{\hat \xi}_j}{\rho}} \otimes
 {\hat \pi_j} $.}
$$
x_j \sim {\hat \xi_j}~,~~~t \sim \sum_j {\hat \pi_j}
$$
$$
[\hat \xi_j , \hat \pi_k] = i \delta_{jk} \hbar_{\text{pregeom}} (1-e^{ - \hat
\xi_j / \rho }),\quad[\pi_j,\pi_k]=0 ~.
$$

\section{Introducing stringspaces}

\subsection{Fuzzy spacetime points with noncommutativity and extendedness}
Also as an illustrative example of how the new perspective on pregeometry
here proposed might be exploited,
in this section we introduce a new candidate
for the structure of spacetime, which would not be naturally contemplated by using
intuitive reasoning directly at the level of spacetime itself, but is indeed a rather
natural candidate from our pregeometric perspective.
We have discussed in the previous sections how the Moyal plane and other noncommutative
spaces (and spacetimes) can be described in terms of the phase spaces of particles "living"
in a classical pregeometric/fictitious spacetime.
With so much to be learned on possible spacetime geometries by studying the
phase spaces of particles it is indeed natural to wonder whether
something meaningful/useful could be found by exploring the possibility
to describe spacetime, one way or another, in terms of the phase spaces
of strings "living"
in a classical pregeometric spacetime.

At first this hypothesis might appear to be rather peculiar, because the coordinates of
a string in phase space are fields, and field coordinates do not look like natural candidates
to describe the coordinates of points. However, the physics community has already
well established that our present picture, with point particles propagating in
a spacetime of points, can be meaningfully generalized by allowing for ``string particles"
propagating in a spacetime of points. In particular the string-theory literature
(and by now even string-theory textbooks) shows very clearly how such extended
objects would appear as point particles to all measuring devices enabled with presently-available
sensitivities (but could add new physics in a high-energy ``stringy regime").
It is therefore not inconceivable that a ``deformation" of similar (but different, see below)
type might be achieved by instead describing particles in terms of close-to-ordinary point-particle
fields propagating however in a spacetime
with ``extended points".

While we are unprepared to offer a fully articulated hypothesis of formulation
of such a ``stringspace", we hope that the exploratory formulation and observations
reported in this section will suffice to illustrate the empowerment that can
come from our new perspective on pregeometry, and perhaps also illustrate
some reasons for our stringspaces to be further explored.

Just because our formalization of the idea of stringspaces is at present
very rudimentary it is perhaps useful, even before getting to it, to offer
some intuition for the type of novel features for spacetime geometry
that more refined formulations of stringspaces might end up providing.
The simplest example we can propose is based on the renowned
Amati-Ciafaloni-Veneziano/Gross-Mende Generalized Uncertainty Principle (GUP)
for string theory~\cite{vEacfEgmEkpp}. Investigations of this GUP establish
that the Heisenberg phase-space uncertainty
principle, $\delta X \delta P \ge 1$,
is modified in the string-theory setting in such a way that (in natural units)
\begin{equation}
\delta X  \ge \frac{1}{\delta P} + L_s^2 \delta P
\end{equation}
and as a result $\text{min} (\delta X) \simeq L_s$ (with $L_s$ the string length).
If for these string-theoretic phase-space coordinates $X$ and $P$
we now contemplate, according to our novel perspective,
a pregeometric reinterpretation as coordinates of space we would have a picture
that is somewhat similar to the one of the Moyal plane, but with more stringent
uncertainty principles for the localization of points on the plane:
on the Moyal plane a single coordinate can be established with arbitrarily high accuracy
(at the price of loosing all information on the other coordinate) but on a stringspace whose
pregeometry is governed by the GUP even attempts to determine a single coordinate
would be affected by an unavoidable minimum-uncertainty constraint.
Several arguments based on the quantum-gravity problem
have advocated exactly this type of fuzziness, and perhaps for these
intuitions some suitable formulation of stringspaces might prove to be a valuable
conceptual tool.

\subsection{Exploratory formulation of a two-dimensional stringspace}
While we envision the possibility to introduce some rather rich formulations of stringspaces,
at present we are only able to offer a very rudimentary formulation.
The pregeometric picture of our exploratory formulation of a two-dimensional
stringspace is ``stringy'' only in the sense that it involves a pair of
fields $\phi(\sigma,\tau)$, $\pi(\sigma,\tau)$, defined on a pregeometric world-sheet
of coordinates $\sigma$ and $\tau$ ($0 \leq \sigma \leq \ell$, $- \infty \leq \tau \leq \infty$),
with $\phi(\sigma,\tau)$ playing the role of a coordinate (-field) of the pregeometric string
in the pregeometric target space and $\pi(\sigma,\tau)$ playing the role
of ``momentum field", conjugate to $\phi(\sigma,\tau)$.

In our present rudimentary formulation of the two-dimensional stringspace
the quantization of the pregeometric picture is introduced for (conceptual) simplicity
through the functional
Schroedinger picture, so that quantum states of the fields are codified in pregeometric ``wave"
functionals $\Psi[\phi(\sigma)]$, with evolution in $\tau$ described by~\cite{Jackiw}
\begin{equation}
i\frac{\partial}{\partial \tau}\Psi[\phi]=H[\hat\phi,\hat\pi]~\Psi[\phi]
~,
\end{equation}
in terms of some Hamiltonian $H[\hat\phi,\hat\pi]$,
and with the (pregeometric) ``observables" $\hat \phi (\sigma)$ and $\hat \pi(\sigma)$
acting respectively by simple multiplication and by a functional derivative:
\begin{equation}
\hat \phi (\sigma) \triangleright \Psi[\phi] = \phi (\sigma) \Psi[\phi] ~ ,
\label{actionphi}
\end{equation}
\begin{equation}
\hat \pi (\sigma) \triangleright \Psi[\phi]
= - i \hbar_{\text{pregeom}}  \frac{\delta \Psi[\phi]}{\delta \phi (\sigma) }  ~ .
\label{actionpi}
\end{equation}
These of course ensure that
\begin{equation}
[\hat \phi(\sigma), \hat \pi(\sigma')] = i \hbar_{\text{pregeom}} \delta(\sigma - \sigma') ~ .
\end{equation}
In analogy with what discussed before for the pregeometric picture of the Moyal plane,
we associate to this pregeometric picture a description of the (would-be-physical) space geometry
in which the fundamental entities are ``extended points" with non-commutative (field-)
coordinates $x_1(\sigma)$ and $x_2(\sigma)$ describable
in terms of the phase-space fields of the pregeometric string:
 $x_1(\sigma) \sim {\hat \phi}(\sigma)$ and $x_2(\sigma) \sim {\hat \pi}(\sigma)$.
 But of course\footnote{The extendedness of spacetime points, here conceptualized,
 would have to be mostly undetectable, if at all present, since
 we must of course reproduce the indications of our observations, so far all
 providing no evidence of such an extendedness.}
  the primary characterization of the stringspace should
 be given in terms of $\overline{x_1}$ and $\overline{x_2}$,
\begin{equation}
\overline{x_1} \sim \frac{1}{\ell} \int_0^\ell
d \sigma \hat \phi(\sigma)
\end{equation}
\begin{equation}
\overline{x_2} \sim \frac{1}{\ell} \int_0^\ell
d \sigma \hat \pi(\sigma)
~,
\end{equation}
which essentially give the average values of the fields $x_1(\sigma)$ and $x_2(\sigma)$.

It is straightforward to notice that
\begin{equation}
[\overline{x_1},\overline{x_2}] \sim \frac{1}{\ell^2} \int_0^\ell \int_0^\ell
d \sigma d \sigma' [\hat \phi(\sigma),\hat \pi(\sigma')] = i \frac{\hbar_{\text{pregeom}}}{\ell} ~ ,
\end{equation}
and therefore $\overline{x_1}$,$\overline{x_2}$ (which however codify only part,
the non-extendedness part, of the structure of the two-dimensional stringspace)
are governed by a commutation relation that is exactly of Moyal-plane type (see Eq.~(\ref{moyal})),
with $ \theta = \hbar_{\text{pregeom}}/\ell$.
As a way to stress this point we shall in this section use interchangeably
the notations $\hbar_{\text{pregeom}}/\ell$ and $ \theta$.

\subsection{Some remarks on the distance observable}
This rudimentary formulation of stringspaces, which at present is the best we
are able to offer, still allows us to at least articulate some
issues and some observations that should be relevant for the analysis
of the distance observable even in more structured stringspaces.
Naturally we would like to treat distances in stringspace making profit at least in part
of the simple technology introduced in Section~2 in the simpler context of the Moyal plane.
To some extent we find that this can be done, and a clear way to illustrate this point
is offered by the simplicity on a special class of Schroedinger-picture functionals
in the pregeometric picture, the  ``gaussian wavefunctionals",
\begin{equation}
\Psi_\Omega[\phi] = {\det}^{-\frac{1}{4}} \left[ \frac{\Omega}{\pi} \right] \exp \left( \int
d \sigma d \sigma' \phi(\sigma) \Omega(\sigma,\sigma') \phi(\sigma')  \right) ~ ,
\end{equation}
which usually play a rather special role in the Schroedinger functional picture\footnote{In particular,
one can form Fock bases starting
from them and applying creation operators \cite{Jackiw}.}.

We shall work with them assuming that the function $\Omega(\sigma,\sigma')$
(the ``covariance" of the gaussian functional)
is real and positive. And the reader should also
notice that our functional-determinant pre-factor ${\det}^{-\frac{1}{4}} \left[ \frac{\Omega}{\pi} \right] $
ensures normalization $\int \mathcal{D}[\phi] \Psi_\Omega^*[\phi]\Psi_\Omega[\phi] = 1$.

These Schroedinger-picture gaussian functionals are rather obviously of interest
for us since they could be to stringspaces what the ``pregeometric
gaussian wavefunctions" are to the Moyal plane: these might contain the pregeometric
description of the quantum state of the geometry of spacetime in which the distance
between two points is minimal.

A proper full
characterization of the concept of distance between two (extendedness-endowed)
points of stringspace, of (field-) coordinates $\{ x_1^{(1)}(\sigma) , x_2^{(1)}(\sigma) \}$
and $\{ x_1^{(2)}(\sigma) , x_2^{(2)}(\sigma) \}$,
 should probably be multilayered, but clearly a key first level
of characterization should be describable in terms
of $\{ \overline{x_1^{(1)}} , \overline{x_2^{(1)}} \}$
and $\{ \overline{x_1^{(2)}} , \overline{x_2^{(2)}} \}$.
And for this level of characterization
we do have a natural candidate for the distance observable,
which is of course
based on the correspondence between $\{ \overline{x_1} , \overline{x_2} \}$
and coordinates in the Moyal plane
discussed at the end of the previous section:
\begin{eqnarray}
{\overline{ d}^2} &\equiv& \left(\overline{x_1^{(1)}}-\overline{x_1^{(2)}} \right)^2
+\left(\overline{x_2^{(1)}}-\overline{x_2^{(2)}}\right)^2 ~.
\label{dsquared}
\end{eqnarray}

We shall of course characterize the properties of ${\overline{ d}^2}$ pregeometrically,
in terms of properties of the corresponding phase-space observables acting
on states of two strings (one string for each point in stringspace that appears
in the geometric spacetime observable, just like in the case of the Moyal plane
two pregeometric particles intervened in the description of geometric
observables for pairs of points of the Moyal plane).
Let us first investigate the properties of ${\overline {d}^2}$
in the case in which our gaussian wave functionals have the
following  simple form
\begin{equation}
\Psi_\omega [\phi_1 , \phi_2]= {\det}^{-\frac{1}{2}} \left[ \frac{\omega \delta}{\pi} \right] \exp \left[ \int
d \sigma  ( {\phi_1}^2(\sigma) +{\phi_2}^2(\sigma) )  \omega(\sigma)\right] ~,
\end{equation}
which essentially amounts to placing each of the two pregeometric strings
in a state described by a gaussian functional with covariance function $\Omega$
of the special form $\Omega(\sigma,\sigma') = \omega(\sigma) \delta (\sigma-\sigma')$
(with positive $\omega(\sigma)$, as a result of the assumed positivity of $\Omega(\sigma,\sigma')$).

In order to evaluate ${ \overline{d}^2}|_{\Psi_\omega}$ (which is our notation for
the expected value of ${ \overline{d}^2}$ in a state of the geometry of stringspace whose pregeometric
description is given by $\Psi_\omega [\phi_1 , \phi_2]$)  it is useful to first notice,
using (\ref{actionphi}),
(\ref{actionpi})
and elementary steps of functional integration and functional differentiation,
that
\begin{equation}
\left< \Psi_\omega \right| \hat d^2 (\sigma,\sigma') \left| \Psi_\omega \right>
= \int \mathcal{D}[\phi] \Psi_\omega^* [\phi] \hat d^2 (\sigma,\sigma') \Psi_\omega [\phi]
= \left( \frac{1}{\omega (\sigma)} +  \ell^2 \theta^2 \omega (\sigma) \right) \delta(\sigma - \sigma')
~ ,
\label{dsquaresigmasigma}
\end{equation}
for
\begin{eqnarray}
\hat d^2(\sigma,\sigma') &\equiv& \left[ \hat \phi_1(\sigma)
- \hat \phi_2(\sigma) \right]\left[ \hat \phi_1(\sigma')- \hat \phi_2(\sigma') \right]
 + \left[ \hat \pi_1(\sigma)- \hat \pi_2(\sigma) \right]\left[ \hat \pi_1(\sigma')
- \hat \pi_2(\sigma') \right] ~ .~~~~~~~~
\end{eqnarray}
From the result (\ref{dsquaresigmasigma})
one then easily concludes that
\begin{equation}
{ \overline{d}^2}|_{\Psi_\omega} =
\frac{1}{\ell^2}\int_0^\ell \int_0^\ell d\sigma  d\sigma'
[\left< \Psi_\omega \right| \hat d^2 (\sigma,\sigma') \left| \Psi_\omega \right>]
= \int d \sigma \left( \frac{1}{ \ell^2 \omega (\sigma)} +  \theta^2 \omega (\sigma) \right) ~ .
\end{equation}

Interestingly this (taking into account the positivity of $\omega(\sigma)$)
allows us to conclude that, at least when restricted on states of
the type $\Psi_\omega$, the expectation values of ${ \overline{d}^2}$ are never smaller
than $2 \theta$: $\text{min}({ \overline{d}^2}|_{\Psi_\omega})=2\theta$.
We conjecture that it should be possible to verify (or easily find a relatively
minor reformulation such) that on our rudimentary two-dimensional stringspace this property
holds for all admissible states and not only for the restricted class $\Psi_\omega$:
\begin{equation}
{ \overline{d}^2}|_{\Psi} \geq 2 \theta ~~~~~~ \forall \Psi
~.
\end{equation}
And among the states ${\Psi_\omega}$ we do find an ``eingenstate of ${ \overline{d}^2}$"
that saturates the bound. This is the following state
\begin{equation}
\Psi_{\omega_{\text{min}}} = {\det}^{-\frac{1}{2}}
 \left[ \frac{ \delta}{\pi \ell \theta} \right] \exp \left[\frac{1}{\ell\theta} \int
d \sigma  ( {\phi_1}^2(\sigma) +{\phi_2}^2(\sigma) ) \right],
\end{equation}
which actually is such that
\begin{equation}
\hat d^2(\sigma,\sigma')\Psi_{\omega_{\text{min}}}[\phi_1,\phi_2]
=2 \theta\ell \delta(\sigma-\sigma')\Psi_{\omega_{\text{min}}}[\phi_1,\phi_2] ~.
\label{thisresult1}
\end{equation}
In light of this result (\ref{thisresult1}) it is then evident that
\begin{equation}
\frac{1}{\ell^2}\int_0^\ell\int_0^\ell d\sigma d\sigma'\left\{\hat d^2(\sigma,\sigma')\Psi_{\omega_{\text{min}}}[\phi_1,\phi_2]\right\}
=2 \theta \Psi_{\omega_{\text{min}}}[\phi_1,\phi_2]
\end{equation}
which indeed amounts to the fact that $\overline{d}^2=2\theta$ sharply
in the state described pregeometrically by $\Psi_{\omega_{\text{min}}}$.

Other noteworthy properties of $\Psi_{\omega_{\text{min}}}$ that are easily
verified are
\begin{equation}
(\overline{x_1}^{(1)}-\overline{x_1}^{(2)})|_{\Psi_\omega}=0
=(\overline{x_2}^{(1)}-\overline{x_2}^{(2)})|_{\Psi_\omega}
\label{joc12a}
\end{equation}
and
\begin{eqnarray}
\int_0^\ell d \sigma \int_0^\ell d \sigma' \left< \Psi_{\omega_{\text{min}}}  \right|
 &&\!\!\!\! [\hat \phi_1(\sigma)
- \hat \phi_2(\sigma)] [\hat \phi_1(\sigma')  - \hat \phi_2(\sigma')] \left|
 \Psi_{\omega_{\text{min}}} \right> = \label{joc12b}\\
 && = \theta  =
\int_0^\ell d \sigma \int_0^\ell d \sigma' \left< \Psi_{\omega_{\text{min}}}  \right| [\hat \pi_1(\sigma)
 - \hat \pi_2(\sigma)][\hat \pi_1(\sigma') - \hat \pi_2(\sigma')] \left| \Psi_{\omega_{\text{min}}} \right>
~.
\nonumber
\end{eqnarray}
In particular this result (\ref{joc12b}),
in light of (\ref{joc12a}),
should be viewed as the pregeometric characterization of the fact that
both $\overline{x_1}^{(1)}-\overline{x_1}^{(2)}$ and $\overline{x_2}^{(1)}-\overline{x_2}^{(2)}$
have uncertainty $\sqrt{\theta}$ when the
pregeometric state is $\Psi_{\omega_{\text{min}}}$.
This may invite a characterization of $\Psi_{\omega_{\text{min}}}$ as a minimum-uncertainty
(pregeometric) state.

In closing this subsection exploring the concept of distance in a two-dimensional stringspace
we find appropriate to stress that,
while the study of ${\overline d^2}$ provides the most intelligible (because most familiar looking)
characterization of distances in (our rudimentary) two-dimensional stringspace,
it exposes only marginally the implications of the ``extendedness" of the points of stringspace.
The little insight we are gaining in
our preliminary (and presently unpublishable) investigations of this extendedness
appears to suggest that its meaningful characterization may prove to
be a formidable challenge. The only tangible contribution we can presently offer
is a sort of ``probe" of this fuzziness, which we characterize through the action of the
following pregeometric-phase-space observable
\begin{equation}
{\hat { \cal D}}^2 \equiv \frac{1}{\ell} \int_0^\ell \int_0^\ell d \sigma
d \sigma' \hat d^2 (\sigma,\sigma') \delta(\sigma-\sigma') ~ ,
\end{equation}
on the following gaussian wave functional
\begin{eqnarray}
&& \Psi_\Omega[\phi_1,\phi_2] = \Psi_\Omega[\phi_1]\Psi_\Omega[\phi_2] = \nonumber \\
&& {\det}^{-\frac{1}{2}} \left[ \frac{\Omega}{\pi} \right] \exp \left( \int
d \sigma d \sigma' \left\{ \phi_1(\sigma) \Omega(\sigma,\sigma') \phi_1(\sigma')
  + \phi_2(\sigma) \Omega(\sigma,\sigma') \phi_2(\sigma')  \right\} \right) ~ .
\end{eqnarray}

In order to compute the expectation of ${\hat { \cal D}}^2$ in $\Psi_\Omega[\phi_1,\phi_2]$
it is useful to first observe that\footnote{Eq.~(\ref{dSguercio}) can be verified with
calculations that are not much different from the ones needed for Eq.~(\ref{dsquaresigmasigma}),
although slightly more tedious.}
\begin{equation}
\left< \Psi_\Omega[\phi_1,\phi_2] \right| \hat d^2 (\sigma,\sigma') \left| \Psi_\Omega[\phi_1,\phi_2] \right>
= \Omega^{-1}(\sigma , \sigma') +  \ell^2 \theta^2 \Omega(\sigma , \sigma') ~ ,
\label{dSguercio}
\end{equation}
where $\Omega^{-1}(\sigma , \sigma')$ is the inverse kernel
of the covariance, defined by $\int d \sigma' \Omega^{-1}(\sigma , \sigma') \Omega(\sigma' , \sigma'')
= \int d \sigma' \Omega(\sigma , \sigma') \Omega^{-1}(\sigma' , \sigma'')
= \delta (\sigma - \sigma'') $.

On the basis of (\ref{dSguercio}) it is then easy to conclude that
\begin{eqnarray}
&& \left< \Psi_\Omega[\phi_1,\phi_2] \right|  \hat { \cal D}^2  \left| \Psi_\Omega[\phi_1,\phi_2] \right>
= \frac{1}{\ell} \int_0^\ell d \sigma \left[ \Omega^{-1}(\sigma , \sigma)
+  \ell^2 \theta^2 \Omega(\sigma,\sigma) \right] ~ ,
\label{joc13a}
\end{eqnarray}
and this can be interestingly compared with ${ \overline{d}^2}|_{\Psi_\Omega}$,
for which one easily finds
\begin{eqnarray}
{ \overline{d}^2}|_{\Psi_\Omega}=
\frac{1}{\ell^2}\int_0^\ell \int_0^\ell d\sigma  d\sigma' [\left< \Psi_\Omega \right|
 \hat d^2 (\sigma,\sigma') \left| \Psi_\Omega \right>]
=  \frac{1}{\ell^2} \int_0^\ell \int_0^\ell d\sigma d \sigma' \left[
\Omega^{-1}(\sigma,\sigma') + \ell^2 \theta^2 \Omega(\sigma,\sigma') \right]
~ .~~~~~~~~~~
\label{joc13b}
\end{eqnarray}
By contemplating both (\ref{joc13a}) and (\ref{joc13b}) one can gain, in spite of
their limited role in the overall characterization of a stringspace, some valuable
insight. The most important point is that whereas $\overline{d}^2$
can be expressed (see (\ref{dsquared})) in terms of
 the ``average coordinates", $\overline{x_1^{(1)}}$, $\overline{x_1^{(2)}}$,
  $\overline{x_2^{(1)}}$,$\overline{x_2^{(2)}}$,
no association with the average coordinates is possible for $\hat { \cal D}^2$.
The average coordinates are the {\it trait d'union} between the Moyal plane
and our two-dimensional stringspace: the stringspace has much more structure then
the Moyal plane but if we restricted our investigations of stringspace only
to entities describable in terms of the average coordinates (which, as stressed above,
satisfy a standard
Moyal-algebra commutator) the differences between our two-dimensional stringspace
and the Moyal plane would probably not be very significant.
The features of stringspace that are more novel are probed by tools
like $\hat { \cal D}^2$, which have no Moyal-plane counterpart, and therefore
are likely to characterize the additional fuzziness\footnote{The Moyal plane
is already ``fuzzy" in the sense of ordinary spacetime noncommutativity,
but on top of that our two-dimension stringspace has other sources of fuzziness
associated with its peculiar feature of having points endowed with extendedness.}
of stringspace.

\subsection{Some candidate additional structures for stringspaces}
The concept of stringspace is potentially rather broad: a space whose points
are described in terms of the (possibly functional) positions in phase space
of some strings ``living" in an
auxiliary spacetime can be many things, reflecting the many possible
ways to characterize the pregeometric strings.
The rudimentary exploratory description of a two-dimensional stringspace
we offered in the previous two subsections is
formulated at a level such that essentially one only
takes into account a few features of
the kinematical Hilbert space~\cite{leeKINEM,carloKINEM,perezreview}
 of the pregeometric string.
For some of the most interesting spacetime features that one might envisage introducing
through a stringspace construction it might be beneficial to enrich the pregeometric
picture with at least some of the structures that a full string theory in the pregeometry
could provide. This in particular appears to be necessary in order to realize
the ``spacetime GUP", which we discussed in Subsection 5.1 as one of the most
appealing opportunities for stringspace modelling, since the
derivations~\cite{vEacfEgmEkpp} of the string-theoretic phase-space GUP
appear to require a fully structured string theory. What exactly would be strictly
needed as structure of the pregeometric string theory in order to get the GUP stringspace
is not completely clear, as a result of the fact that the role of the standard GUP
in the foundational conceptual ingredients of string theory is still not well
understood~\cite{wittenPhysicsToday1995}.
But in general (and perhaps in particular for this GUP-stringspace issue)
it appears to us, consistently with the observation we reported in Section~4,
that one should not exclude {\it a priori}
the possibility of a fully dynamical formulation of the pregeometry, and that
one should at least
contemplate the possibility to associate some kinematical/structural aspects
of the stringspace to genuinely dynamical entities of the pregeometric level.

Another key issue which was not considered in the preliminary exploration of stringspaces
reported in this section is the one of symmetries. In a stringspace setup there are at least three
levels at which symmetry issues should be contemplated: symmetries of the pregeometric
world-sheet of the string, symmetries of the pregeometric target space of the string,
and, most importantly, symmetries of the phase space of the pregeometric string, which
will affect directly the symmetry structure of the stringspace.
It is perhaps amusing to observe that, even when the stringspace is intended
as a nonclassical description of a quasi-Minkowski spacetime,
one might well have a valuable formulation
of the stringspace in cases in which the pregeometric target space is not Poincar\'e invariant,
not even approximately. In such instances one would have to insist
on (at least approximate) Poincar\'e symmetry
of the stringspace, but this translates into a demand for the pregeometric phase space
and not necessarily the pregeometric target space.
In some cases however this might amount to a rather involved technical issues
for the level of the pregeometric phase space, as one would notice immediately
in the analysis of stringspaces of 3 or more dimensions.
In fact, while the Moyal plane is essentially trivial~\cite{agmNEXT} from a symmetry perspective
(and at least some of that triviality should be reflected in the pregeometric phase space),
the generalizations to 3 or more spacetime dimension (see {\it e.g.} our Section~4)
are often described in terms of Hopf-algebra
symmetries~\cite{lukieold,majidruegg,lukieAnnPhys,chaichan,wess,bala,thetanoether}
and stringspaces that in the ``average coordinate limit" resemble such
noncommutative spacetimes may well be affected by this feature.

Of course, if stringspaces were ever to be used to do physics,
we would not want just theories {\underline{of}} stringspace but actual theories
formulated {\underline{in}}
stringspace. And in this respect it should be noticed that, for example, a stringspace
version of an ordinary theory
of ``classical fields in commutative Riemannian spacetime" in our rudimentary two-dimensional
stringspace would have to involve functionals of noncommutative (quantum) fields. For example,
a metric field should be of the
type $G_{ij}[x_1(\sigma),x_2(\sigma)] \sim G_{ij}[{\hat \phi}(\sigma),{\hat \pi}(\sigma)]$.
It is worth noticing how different this is from the D-dimensional target-space metric field
that one introduces in writing certain actions in the standard string-theory framework, which
is of the type $G_{\mu,\nu}[\phi_1(\sigma,\tau),\phi_2(\sigma,\tau),...,\phi_{D}(\sigma,\tau)]$.
In particular $G_{ij}[{\hat \phi}(\sigma),{\hat \pi}(\sigma)]$ must reflect the noncommutativity
between ${\hat \phi}(\sigma)$ and ${\hat \pi}(\sigma)$,
whereas $G_{\mu,\nu}[\phi_1(\sigma,\tau),\phi_2(\sigma,\tau),...,\phi_{D}(\sigma,\tau)]$
depends exclusively on the $\phi_\mu$ coordinate fields on the world-sheet
which of course all commute with one another.
It is also worth stressing
that in the study of theories in stringspace a natural fourth level of symmetry analysis
would arise
(in addition to the three natural levels of symmetry analysis mentioned above),
since one would need to establish the symmetries of these theories introduced in stringspace.

In closing this section, after mentioning so many structures
that could (and probably should) be added to our pregeometric-string picture,
let us also mention that we might however have introduced at least one more feature
than needed: guided by the analogy with the use of pregeometry in
the study of noncommutative spacetimes we immediately assumed that the pregeometric
string should be quantized, rendering its phase space noncommutative,
so that indeed the stringspace coordinates would
be described by noncommutative quantum fields. It is however at least not obvious that
one should necessarily proceed in this way: classical pregeometric strings
may well lead to meaningful scenarios, some ``commutative stringspaces",
and probably non-trivial scenarios, since,
even without the noncommutativity feature, the corresponding stringspaces
would still be affected by fuzziness induced by the extendedness provided
to the points.
And it is amusing to wonder whether there could be a
a theory appropriately formulated in a ``commutative stringspace" that would
reproduce string theory:
it is  not inconceivable that in some specific setups the extendedness
of particles (replacing a particle by a string) could be traded for an extendedness of points
(replacing classical Minkowski space with a Minkowskian commutative stringspace).

\section{Outlook}
The techniques used in our Sections~2 and 3, within a
conventional setup of exploitation of the pregeometry perspective,
should be applicable to a rather broad collection of geometric observables
of the Moyal plane, and of other noncommutative spacetimes.
Our observations on the area observable (which shall be extended in Ref.~\cite{agmNEXT})
and on the distance observable appear to suggest that there is no universal behaviour
of this type of observables: for distances we derived the presence of a minimum nonzero value,
while zero area is possible. And this (as stressed in Section~3)
appears to reflect correctly what one should expect
on the basis of the form of the Moyal-plane commutator of coordinates.
Besides providing a meaningful characterization of Moyal-plane noncommutativity,
this observation should prompt a careful reexamination of some of the
semi-heuristic arguments that appeared in the quantum-gravity literature,
especially during the last decade. Several of these arguments motivate rather rigorously
(as rigorously as possible using heuristic arguments) the existence
of a minimum-length bound, and then with somewhat loose use of logics assume
that one would reach similar conclusions for areas and volumes, as if in a quantum
spacetime the presence of a minimum-length bound would necessarily imply corresponding
bounds for areas and volumes. We believe that the Moyal plane, in spite of its relative
simplicit, may provide a significant counter-example for the assumptions
that guide this type of reasoning.

For what concerns
our proposal of string-theoretic pregeometries, and the associated ``stringspaces",
at present the only natural target which appears to be in sight is the
implementation of the ``GUP in spacetime" (here described in Subsection~5.1),
possibly exploiting some of the observations we offered in Subsection~5.4.
If that is accomplished then one could attempt to formulate quantum field theories
in the relevant stringspace, and in particular we conjecture that such
quantum field theories would be immune from the peculiar ``IR/UV mixing"~\cite{iruv1,iruv2,iruvgac}
which instead appears to affect quantum field theories in Moyal/canonical noncommutative
spaces. IR/UV mixing is a result of the connection between short distances in one direction
and large distances in another direction which are generated by Moyal-type uncertainty
relations (the type $\delta x_1 \delta x_2 \geq \theta$) and we believe it
should not be present if instead the uncertainty relations are such that for all coordinates $x_i$
one has $\delta x_i \geq \sqrt{\theta}$ (which is what we expect of the ``GUP in spacetime").
But our proposal of stringspaces is perhaps even more interesting as an illustrative
example of a novel way to make use of the pregeometry perspective (here advocated
in Section~4) which transforms the whole ensemble of results of theoretical
physics into a reservoir of ideas/intuitions for what could be the structure
of spacetime at short (possibly Planckian) distances.

\section*{Acknowledgments}
G.~A.-C. is supported by grant RFP2-08-02 from The Foundational Questions Institute (fqxi.org).\\
G.~G. is supported by ASI contract I/016/07/0 "COFIS".

\end{document}